# Does the evolution of complex life depend on the stellar spectral energy distribution?

Jacob Haqq-Misra
Blue Marble Space Institute of Science
1001 4th Ave, Suite 3201, Seattle, Washington 98154, USA
Email: jacob@bmsis.org

**Abstract**

This paper presents the proportional evolutionary time hypothesis, which posits that the mean time required for the evolution of complex life is a function of stellar mass. The "biological available window" is defined as the region of a stellar spectrum between 200 to 1200 nm that generates free energy for life. Over the ~4 Gyr history of Earth, the total energy incident at the top of the atmosphere and within the biological available window is $\sim 10^{34}$ J. The hypothesis assumes that the rate of evolution from the origin of life to complex life is proportional to this total energy, which would suggest that planets orbiting other stars should not show signs of complex life if the total energy incident on the planet is below this energy threshold. The proportional evolutionary time hypothesis predicts that late K- and M-dwarf stars ($M < 0.7 \, M_\odot$) are too young to host any complex life at the present age of the universe. F-, G-, and early K-dwarf stars ($M > 0.7 \, M_\odot$) represent the best targets for the next generation of space telescopes to search for spectroscopic biosignatures indicative of complex life.

## 1. Introduction

Earth serves as the only example of an inhabited planet, so the rate of biological evolution across Earth's history can provide a basis for astrobiological speculation on the evolution of life on other planets. The geologic record of microbial life extends as far back as 3.7 Ga (Pearce et al. 2018), but the subsequent development of complex and intelligent life required another several billion years. This terrestrial history suggests the possibility that other habitable planets may follow similar trajectories, with the rapid emergence of a microscopic biosphere followed by a prolonged period of evolutionary development for complex life. The concept of "complex life" in this context refers to any organisms comparable to multicellular fungi, plants, or animals (Ward and Brownlee 2000).

One approach to estimating the mean time required for the evolution of complex life, $t_{evo}$, is to assume that evolutionary time scales on Earth are typical for other inhabited planets. Haqq-Misra and Kopparapu (2018) refer to this approach as the "equal evolutionary time" (EET) hypothesis. EET assumes that the timing of evolution on Earth is an average case for evolution in general, which suggests that $t_{evo}$ is the mean or expected value for the evolution of complex life on other planets.[1] For example, Lineweaver et al. (2004) follow this assumption of $t_{evo} = 4 \pm 1$ Gyr in order to constrain the spatial and temporal distribution of complex life in the galaxy, while acknowledging that such an assumption is "highly speculative." Catling et al. (2005) similarly suggest that the ∼3.9 Gyr time for atmospheric oxygen to accumulate to levels conducive for complex life on Earth might be applicable to Earth-like planets orbiting other stars. Another way to arrive at EET is through the molecular clock hypothesis, which suggests that the rate at which beneficial genetic mutations accumulate has remained relatively constant across Earth's history (Bromham and Penny 2003). If the rate of evolutionary change on Earth is an average case for evolutionary change on other inhabited planets, then EET suggests that the transition from the origin of life to complex life elsewhere should require approximately the same number of mutations (and thus the same amount of time) on average.

Without any further examples of inhabited planets, EET remains a popular, implicit, or default assumption among many astrobiologists. But the timing of evolution could also depend on interactions between the planet's environment and host star, which may limit the applicability of EET to planets orbiting non-solar type stars. As an alternative to EET, Haqq-Misra and Kopparapu (2018) propose the "proportional evolutionary time" (PET) hypothesis, which suggests that $t_{evo}$ is proportional to the main sequence lifetime of the planet's host star. PET implies that complex life should develop more slowly on planets orbiting low-mass compared to sun-like stars. For example, if the accumulation of abundant atmospheric oxygen is a requirement for the evolution of complex life (Catling et al. 2005), then the timing of oxygenation will likely depend upon the free energy on the planet available for photochemistry and greenhouse warming. Oxygen accumulation is also coupled with other biogeochemical cycles (Falkowski and Godfrey 2008), which in turn may hold functional dependencies on the incoming stellar energy distribution. Likewise, the properties of photosynthetic pigments on other planets may depend upon the energy distribution of the host star (Kiang et al. 2007a,b), which would limit the energy available for oxygenic photosynthesis for planets orbiting low-mass stars (Lemer et al. 2018). Another factor is the availability of ultraviolet radiation from a planet's host star, which may make some stellar spectral types more conducive to beneficial mutations than others (Rugheimer et al. 2015; Ranjan and Sasselov 2016, 2017; Ranjan et al. 2017; Rimmer et al. 2018). Other articulations of the PET hypothesis discuss possible factors that would lead to differences in the timing of evolution based upon stellar spectral type (e.g., Loeb et al. 2016; Ramjan et al. 2017; Haqq-Misra et al. 2018; Lingam and Loeb 2018a,b; Lehmer et al. 2018).

Livio and Kopelman (1990) develop a qualitative relationship between stellar mass and the evolution of complex life by suggesting that $t_{evo}$ depends upon the intensity of ultraviolet radiation incident on the planet. Using simplified mass-luminosity and mass-radius relationships, Livio and Kopelman (1990) calculate the attenuation of ultraviolet radiation from 100 to 200 nm in Earth-like atmospheres across a range of host stars. Ultraviolet radiation is chosen as a proxy for habitability in these calculations due to its utility in demonstrating a functional relationship between the timescale of biological evolution and the lifetime of the host star. In this context, the purpose of Livio and Kopelman's (1990) argument is to provide an alternatives to "statement[s] of Carter and Barrow and Tipler that there is no physical connection between the timescales of stellar and biological evolutions." Livio and Kopelman (1990) argue that the evolutionary timescale on Earth is a monotonic function of stellar mass, such that low-mass stars may be unable to support life at the present time; however, the authors state that further quantification of this transition "is a question that depends on the detailed physics and biology of the development of life."

---

[1] Anthropic reasoning by the self-sampling assumption (SSA) is the one of the most articulate ways to defend EET. SSA states that "one should reason as if one were a random sample from the set of all observers in one's reference class" (Bostrom 2002), where the reference class in this case refers to the set of all planets inhabited with complex life (Haqq-Misra et al. 2018).

This paper develops a formal articulation of the PET hypothesis based upon the total "biologically available energy" over the main sequence lifetime of the host star. The biologically available energy from Earth's history provides a basis for predicting $t_{evo}$ as a function of stellar mass, which suggests that complex life may be less common on planets orbiting low-mass stars at the current age of the universe. This quantitative concept extends the qualitative approach of Livio and Kopelman (1990) by applying stellar evolutionary models to calculate the threshold at which low-mass stars are unlikely to host complex life. Although this method necessarily simplifies the complexities of physics, biology, geology, and other contributors, such an approach provides a first-order estimate of a functional dependence between evolutionary time and stellar mass. The PET hypothesis can be tested with the next generation of space telescopes by searching for spectroscopic signatures from exoplanets that indicate the presence of complex life.

## 2. Biologically Available Energy

From a thermodynamic perspective, a terrestrial planet like Earth acts as a heat engine with incident (low-entropy) stellar radiation providing the primary forcing[2] and outgoing (high-entropy) infrared radiation emitted as waste to the cooler reservoir of space (Hasselman 1976). This planetary heat engine acts to generate free energy that can perform work within the system, which includes geophysical, chemical, and biological processes. Frank et al. (2017) distinguishes four forms of free energy that derive from absorption of incident stellar radiation. Climate circulation and other geophysical motion generates mechanical free energy, $E_{geo}$, while photochemistry and geochemistry generates chemical free energy, $E_{chem}$. Biological processes like photosynthesis generate chemical free energy $E_{life}$, which requires the availability of incident radiation as well as sufficient nutrients provided by $E_{geo}$ and $E_{chem}$. Finally, technological processes such as photovoltaics can generate additional free energy, $E_{tech}$, which directly depends on incident stellar radiation but also draws from $E_{geo}$, $E_{chem}$, and $E_{life}$ for food and resources. The evolution of complex life on Earth has been sustained by direct sunlight as well as by solar-driven mechanical and chemical free energy ($E_{geo}$, $E_{chem}$, $E_{life}$), with modern civilization drawing upon additional solar-driven free energy ($E_{tech}$). This paper assumes that the availability of mechanical and chemical free energy is a limiting factor for the evolution of life on habitable terrestrial exoplanets, with the total free energy ($E_{geo}$ + $E_{chem}$ + $E_{life}$ + $E_{tech}$) bounded by the incident stellar energy.

The first step in articulating the PET hypothesis requires calculating the total incident energy across the history of life on Earth. The solar spectral energy distribution at the top of Earth's atmosphere (Fig. 1, gray curve)[3] corresponds approximately to a blackbody, which is defined by the Planck function,

$$B(T) = \frac{2hc^2}{\lambda^5} \frac{1}{\exp\left(\frac{hc}{\lambda kT}\right)-1}, \tag{1}$$

where $h$ is Planck's constant, $k$ is Boltzmann's constant, $\lambda$ is wavelength, and $c$ is the speed of light. Integrating the radiance in Eq. (1) across the solid angle subtended by the sun at the distance of Earth gives the blackbody irradiance at the top of Earth's atmosphere,

$$F(T) = \frac{\pi R^2}{D^2} B(T), \tag{2}$$

where $R$ is stellar radius and $D$ is the mean distance orbital distance. The blackbody spectral energy distribution, $F(T)$, at a radiating temperature of $T = 5780$ K shows a peak at approximately 500 nm and an integrated irradiance over all wavelengths equal to 1365 W m$^{-2}$ (Fig. 1, black curve).

---

[2] This analysis focuses only on stellar forcing and neglects internal sources such as geothermal energy.

Of this total spectral energy distribution, the PET hypothesis assumes that only photons within a restricted wavelength range are capable of generating the free energy required for complex life. In particular, PET assumes that the free energy available to complex life ($E_{geo} + E_{chem} + E_{life} + E_{tech}$) is linearly[4] proportional $F(T)$, which itself is a function of wavelength. The minimum wavelength, $\lambda_{min}$, is constrained by the limit of shortwave radiation to interact with DNA to cause a mutation or dissociation (thereby limiting $E_{chem}$). This threshold occurs near $\lambda_{min} = 200$ nm, where radiation is capable of interacting with DNA molecules (Buccino et al. 2006; Ranjan et al. 2017; Rimmer et al. 2018; Todd et al. 2018). Radiation at shorter wavelengths than $\lambda_{min}$ may therefore be unusable, and even detrimental, to life. Conversely, ultraviolet radiation at $\lambda_{min}$ and longer wavelengths can benefit life by driving genetic mutations and perhaps even by contributing to abiogenesis (Rugheimer et al. 2015; Ranjan and Sasselov 2016, 2017; Ranjan et al. 2017; Rimmer et al. 2018; Todd et al. 2018). The timing of geologic events, such as the accumulation of atmospheric oxygen (Catling et al. 2005; Falkowski and Godfrey 2008), may also be affected by atmospheric photochemical reactions (i.e, changes in $E_{chem}$) with rates that depend upon photons at wavelengths greater than $\lambda_{min}$. Photosynthesis generates chemical free energy ($E_{life}$) using radiation at wavelengths up to about 800 nm (Kiang et al. 2007a,b), although some anaerobic organisms can use photons up to 1000 nm (Lehmer et al. 2018). Technological life, such present-day human civilization, can generate further free energy ($E_{tech}$) by harnessing even longer wavelength photons using photovoltaics out to near 1200 nm (Blankenship et al. 2011). This maximum wavelength limit, $\lambda_{max} \approx 1200$ nm, is taken to represent the general limit of an incident photon capable of exciting an electron from a medium. Complex life may therefore be limited in its ability to use technology to harness stellar energy at wavelengths much larger than $\lambda_{max}$. This biological available window from $\lambda_{min}$ to $\lambda_{max}$ provides a constraint on the total available flux of "useful" photons for biology and technology.[5] This broad approach to the PET hypothesis enables consideration of the most general sense in which the spectrum of a star could limit the free energy available for driving geophysical, chemical, and biological processes on an orbiting planet.

---

[3] This is the air mass zero reference spectrum developed by the American Society for Testing and Materials (ASTM E-490).

[4] The assumption of a linear monotonic relationship between free energy and the stellar spectral energy distribution is almost certainly false; for example, non-linearities in the climate system or punctuated events like the rise of oxygen would necessitate a non-linear function. However, the linear monotonic assumption is the simplest place to start for the purpose of comparing with the EET hypothesis (which assumes that free energy is wholly unrelated to the stellar spectral energy distribution).

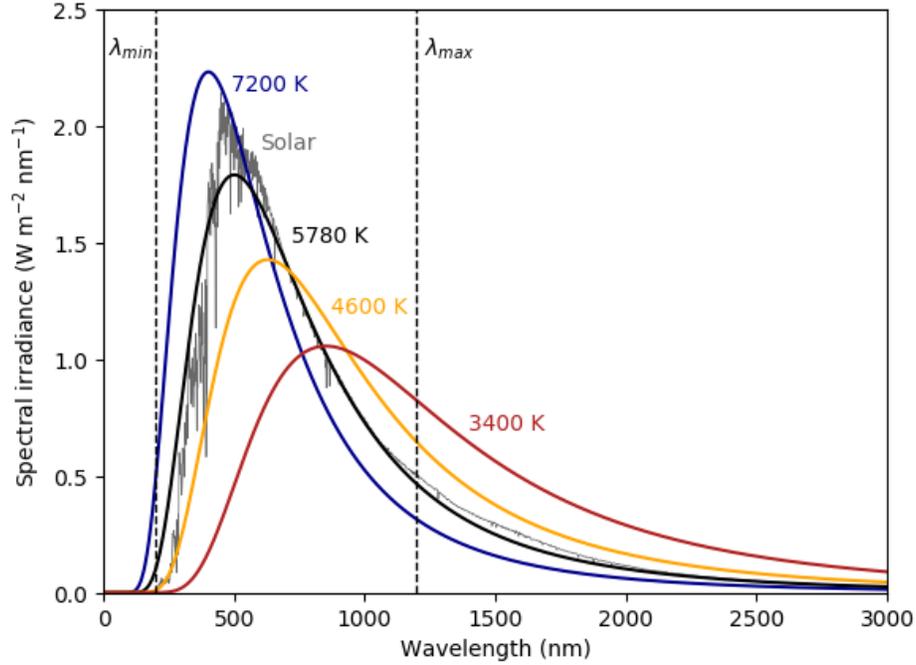

**Figure 1:** The solar spectral energy distribution at the top of Earth's atmosphere (gray) can be approximated by a 5780 K blackbody curve (black). Dashed lines indicate the minimum and maximum wavelength limits for biologically available energy. Blackbody curves for 7200 K (blue), 4600 K (orange), and 3400 K (red) are normalized to the solar constant.

The total power incident on Earth and within the biologically available window is given by integrating the blackbody irradiance, Eq. (2), and multiplying by the cross-sectional area of Earth,

$$P_{bio}(T) = \pi R_\oplus^2 \int_{\lambda_{min}}^{\lambda_{max}} F(T) \, d\lambda, \qquad (3)$$

where $R_\oplus$ is the radius of Earth. The quantity $P_{bio}$ represents the biologically available power from a blackbody spectrum. The biologically available power is $1.39 \times 10^{17}$ W for the 5780 K blackbody and $1.36 \times 10^{17}$ W for the solar spectrum shown in Fig. 1. This value of $P_{bio}$ is approximately 80% of the total solar power incident at the top of Earth's atmosphere. The sun also steadily brightens with time, like all main sequence stars, in response to core contraction from the fusion of hydrogen. The change in luminosity of the sun, $L$, as a function of time, $t$, is expressed by Gough (1981) as,

---

[5] The values of $\lambda_{min}$ to $\lambda_{max}$ can vary by about 20% without significantly changing the results of the PET hypothesis shown by Fig. 3 and Eq. (7).

$$L(t) = L_\odot[1 + \frac{2}{5}(1 - \frac{t}{t_\odot})]^{-1}, \tag{4}$$

where $t_\odot$ = 4.57 Gyr is the age of the solar system and $L_\odot$ is the present-day luminosity of the sun. Eq. (4) describes the change in the total solar power with time across all wavelengths; however, this function can also approximately describe the change in solar power between $\lambda_{min}$ and $\lambda_{max}$. Replacing $L_\odot$ in Eq. (4) with $P_{bio}$ from Eq. (3) and integrating across time gives an expression for the biologically available energy,

$$E_{bio}(T) = \int_0^{t_\odot} P_{bio}(T)[1 + \frac{2}{5}(1 - \frac{t}{t_\odot})]^{-1} dt = \frac{5}{2}\ln(\frac{7}{5})t_\odot P_{bio}(T). \tag{5}$$

Eq. (5) shows that the biologically available energy incident upon Earth across its entire history, for $T$ = 5780 K, is $E_{bio}^\oplus = 1.7 \times 10^{34}$ J.

More sophisticated stellar evolution models provide another method for calculating a value of $E_{bio}$, based upon simulated trajectories of stellar luminosity and effective temperature over the main sequence lifetime (Baraffe et al. 2015). The luminosity of a solar mass star increases steadily during its main sequence phase and is accompanied by modest changes in effective temperature, which (from Eq. (3)) predicts a trajectory for the biological available power (Fig. 2, black curve). Integrating the biological available power from the beginning to present age of the solar system (i.e., the area under the black curve of Fig. 2, bottom panel, from time zero to "now") gives the biological available energy for Earth. This value, $E_{bio}^\oplus = 1.6 \times 10^{34}$ J, remains consistent with the previous estimate from Eq. (5).

The PET hypothesis is expressed in terms of the total energy incident at the top of a planet, rather than the free energy available on its surface. The free energy ($E_{geo} + E_{chem} + E_{life} + E_{tech}$) is indeed the relevant quantity that could cause genetic mutations, drive photochemistry, enable photosynthesis, or otherwise support the evolution of life; however, estimating the free energy would also require knowledge of planetary conditions (such as the planetary albedo or atmospheric composition) over time. Rather than attempt to predict a typical set of planetary parameters or assuming present-day attenuation of radiation (Livio and Kopelman 1990), this articulation of PET treats $E_{bio}^\oplus$ as an upper limit, prior to any attenuation, to the free energy capable of driving biological or technological processes. The purpose of this exercise is to suggest that the quantity $E_{bio}^\oplus$ holds a functional dependence to the stellar spectral energy distribution. Although the EET hypothesis assumes that a timescale of $t_{evo} \approx t_\odot$ is probably typical for complex life to evolve, the PET hypothesis instead suggests that inhabited planets should not develop complex life if the total energy incident upon the planet is less than $E_{bio}^\oplus$.

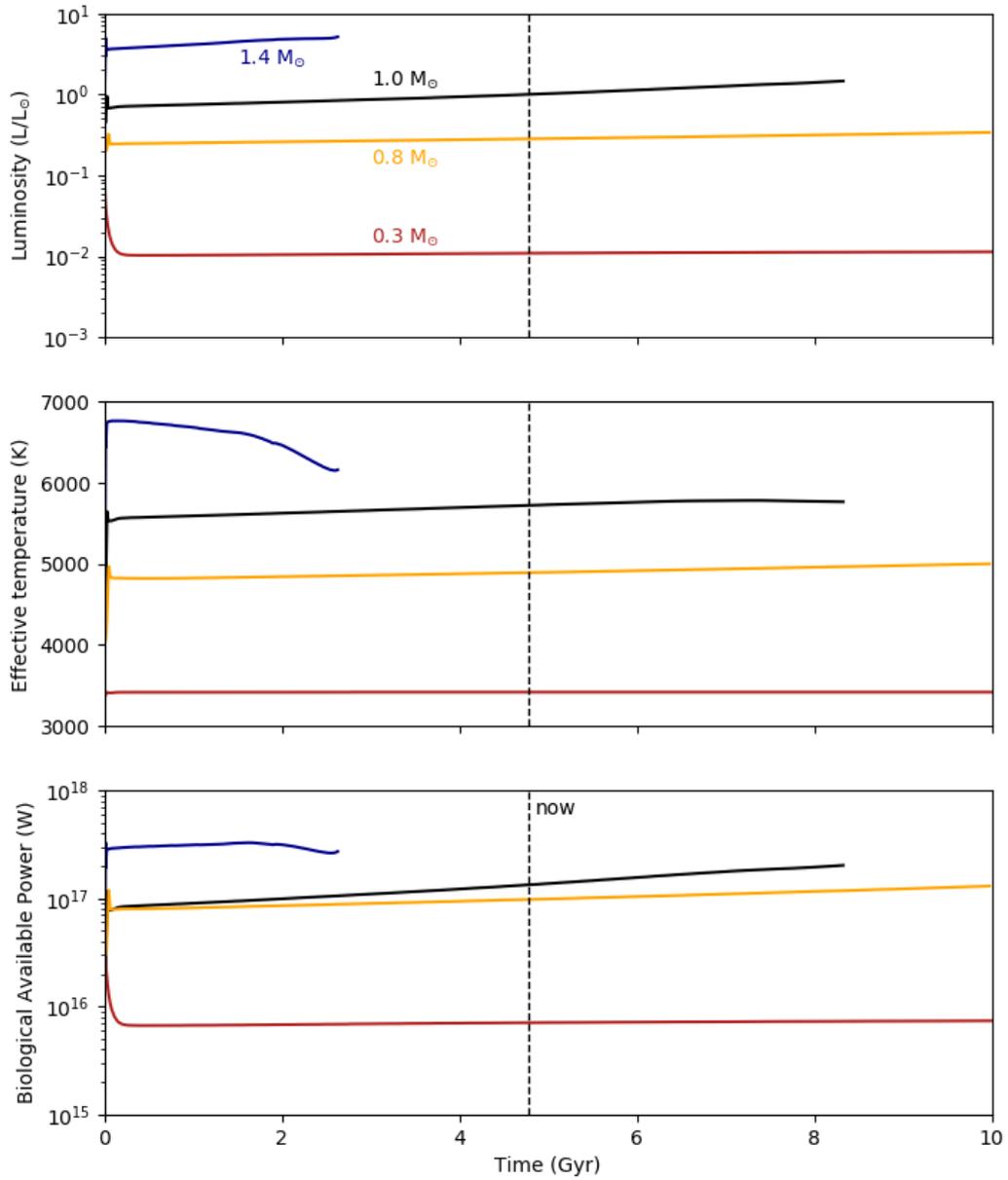

**Figure 2:** The luminosity (top) and effective temperature (middle) from stellar evolution models determine the biological available power (bottom) for 1.4 M$_\odot$ (blue), 1.0 M$_\odot$ (black), 0.8 M$_\odot$ (orange), and 0.3 M$_\odot$ (red) stars. The dashed vertical line indicates the present age of the solar system.

## 3. Evolutionary Timescales for Other Planetary Systems

Fig. 2 shows the biological available power for 1.4 M$_\odot$ (blue curve), 0.8 M$_\odot$ (orange curve), and 0.3 M$_\odot$ (red curve) stars, based upon luminosity and effective temperature trajectories from stellar evolutionary models (Barrafe et al. 2015). The Planck functions[6] for these cases assume a value of $T$ equal to the mean effective temperature over the main sequence. The value of $R$ in Eq. (2) equals the mean stellar radius, and the orbital distance is set to $D = R_\oplus(\langle L \rangle/L_\odot)^{1/2}$ (with $\langle L \rangle$ as the mean luminosity). This value of $D$ represents the "Earth equivalent distance" at which a planet receives the same total stellar power as present-day Earth.[7] This approach demonstrates that even when orbital distance is adjusted to keep total power constant, the biological available power for the 0.3 $M_\odot$ case is still an order of magnitude less than other stars in Fig. 2.

---

[6] The use of Planck functions keeps the expressions analytic, but the same analysis could be applied to empirically measured stellar spectra with similar results.

The quantity $E_{bio}^\oplus$ represents the total energy between $\lambda_{min}$ and $\lambda_{max}$ incident at the top of the atmosphere across all of Earth's history. The PET hypothesis posits that knowing the value of $E_{bio}^\oplus$ enables a general expression for the mean evolutionary time required for complex life, $t_{evo}$, as a function of stellar mass. For any given planetary system, PET states that

$$\int_0^{t_{evo}} P_{bio}(T) \frac{L(t)}{\langle L \rangle} dt = E_{bio}^\oplus . \tag{6}$$

Numerical solutions of Eq. (6) for $t_{evo}$ are plotted in Fig. 3 as filled circles for 1.4 $M_\odot$, 0.8 $M_\odot$, and 0.3 $M_\odot$, with open circles showing additional 1.2 $M_\odot$, 0.6 $M_\odot$, 0.4 $M_\odot$, and 0.1 $M_\odot$ stars (Barrafe et al. 2015). The best fit regression line in Fig. 3 shows a direct relationship between $t_{evo}$ and stellar mass, correlated with $R^2 = 0.96$, and can be expressed as

$$t_{evo} = t_\odot \exp\left[4(1 - \frac{M}{M_\odot})\right]. \tag{7}$$

Eq. (7) predicts that $t_{evo} > 13.8$ Gyr when $M < 0.7$ $M_\odot$. Stars less massive than this limit are too young to host any complex life at the present age of the universe.

Variants of Eq. (7) can be explored by changing the wavelength range between $\lambda_{min}$ and $\lambda_{max}$ to explore other scenarios. Fig. 4 shows the same calculations as in Fig. 3 but with the wavelength range restricted to the photosynthetic range from $\lambda_{min} = 400$ nm to $\lambda_{max} = 700$ nm (left panel) and the ultraviolet between $\lambda_{min} = 200$ nm to $\lambda_{max} = 400$ nm (right panel). Photosynthesis limits the availability of biomass and could therefore be the limiting factor in the development of complex life. Limiting biologically available radiation to the photosynthetic range gives a best fit line of $t_{evo} = t_\odot \exp[5(1 - M/M_\odot)]$, which predicts that $t_{evo} > 13.8$ Gyr when $M < 0.8$ $M_\odot$. Likewise, genetic mutations depend primarily upon ultraviolet radiation and may be the dominant factor leading to the emergence of complex life. Limiting biologically available radiation to ultraviolet only gives a best fit line of $t_{evo} = t_\odot \exp[7(1 - M/M_\odot)]$, which predicts that $t_{evo} > 13.8$ Gyr when $M < 0.9$ $M_\odot$. Fig. 4 shows that the functional dependence between stellar mass and evolutionary timescale remains regardless of the selected wavelength range, although the actual low-mass limit does depend upon the choice of scenario.

---

[7] The Earth equivalent distance does not necessarily correspond to the liquid water habitable zone as calculated by Kopparapu et al. (2013). Using the Earth equivalent distance normalizes the Planck functions for other stellar masses to the solar constant, which enables a consistent comparison of differences in energy across the biological available window.

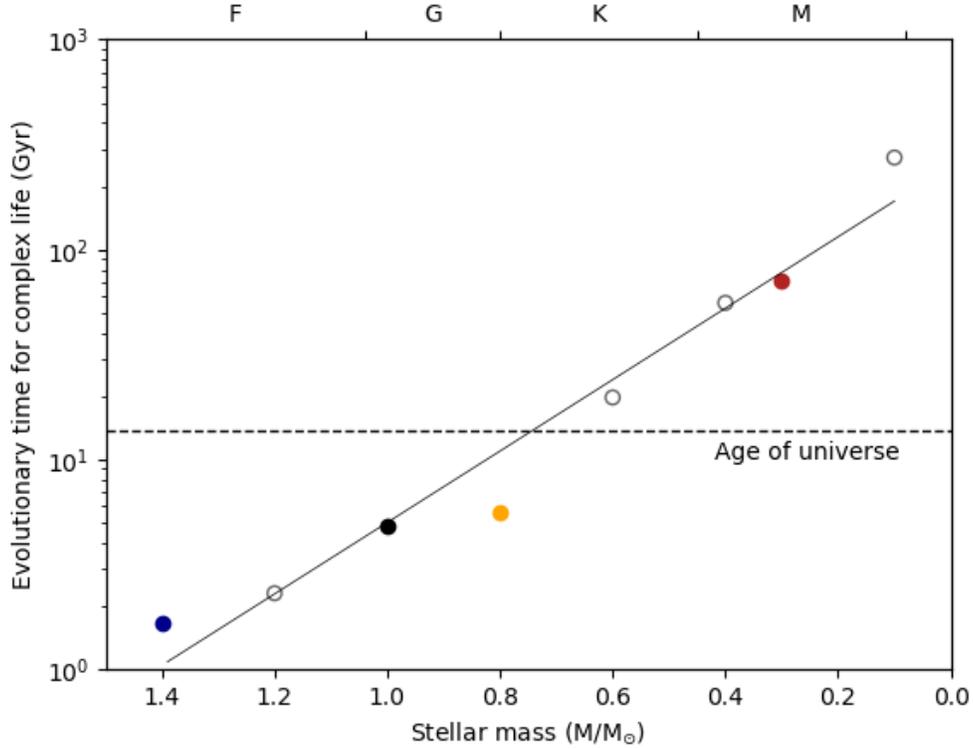

**Figure 3:** The proportional evolutionary time (PET) hypothesis predicts that the mean evolutionary time for complex life depends on the mass of the host star. Filled circles indicate the 1.4 M$_\odot$ (blue), 1.0 M$_\odot$ (black), 0.8 M$_\odot$ (orange), and 0.3 M$_\odot$ (red) stars shown in Fig. 2, while open gray circles show additional 1.2 M$_\odot$, 0.6 M$_\odot$, 0.4 M$_\odot$, and 0.1 M$_\odot$ stars. The top axis indicates the stellar spectral classification, and the dashed horizontal line indicates the present age of the universe. The solid line shows a best fit regression with $R^2 = 0.96$.

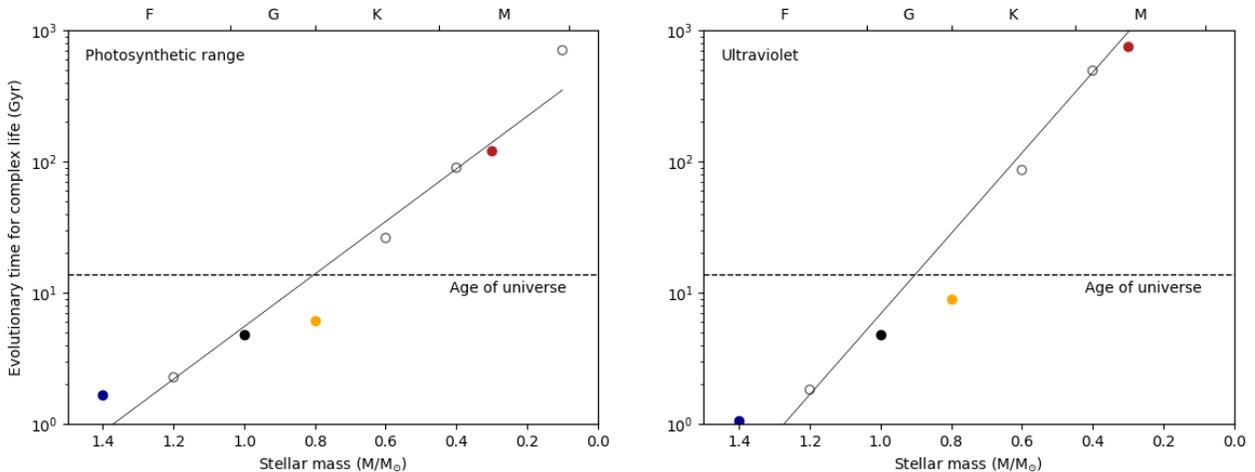

**Figure 4:** Variants of PET show the relationship between the mean evolutionary time for complex life and the mass of the host star when biologically available energy is restricted to the photosynthetic range (left) and ultraviolet only (right). Filled and open circles indicate the stars show in Fig. 2, the top axis indicates the stellar spectral classification, and the dashed horizontal line indicates the present age of the universe. The solid line shows a best fit regression with $R^2 = 0.95$.

## 4. Testing the Proportional Evolutionary Time Hypothesis

The PET hypothesis predicts that G- and early K-dwarf stars ($M > 0.7$ M$_\odot$) are the best targets to search for complex life. Late K- and M-dwarf stars ($M < 0.7$ M$_\odot$) emit limited energy within the biological available window, so any orbiting planets may only develop limited microbial biospheres, if any life evolves at all. PET also predicts that F-dwarf stars could develop complex life after ~2 Gyr. Even though the main sequence lifetime of F-dwarfs is only ~3 Gyr, such stars remain plausible candidates to search for signs of complex life. The PET hypothesis does not imply that complex life is inevitable or even plentiful around F-, G-, and early K-dwarf stars, as a multitude of other factors could limit abiogenesis or otherwise prevent the development of complex life. Instead, this hypothesis suggests that late-K and M-dwarf stars are unlikely to develop complex life until a more distant time in the future. Such a conclusion remains consistent with previous analyses that approach the problem from other perspectives (Loeb et al. 2016; Ramjan et al. 2017; Haqq-Misra et al. 2018; Lingam and Loeb 2018a,b; Lehmer et al. 2018).

The search for spectroscopic biosignatures provides one approach to test this hypothesis. The next generation of space telescopes will enable characterization of nearby planets that will eventually yield a statistically significant sample of spectral signatures from habitable terrestrial planets (Kiang et al. 2018; Fujii et al. 2018). PET predicts that any spectroscopic biosignatures unique to complex life should be absent, or extremely rare, on planets orbiting low-mass stars. Spectroscopic "technosignatures" are the best example of biosignatures exclusive to complex (and intelligent) life, such as the detection of synthetic greenhouse gases like chlorofluorocarbons (Schneider et al. 2010; Lin et al. 2014) or the presence of an absorbing "silicon edge" that indicates the use of planetary-scale technology to harvest stellar energy (Lingham and Loeb 2017). Observable technosignatures could include narrowband radio signals, infrared waste energy, or even Dysonian megastructures (Wright 2018). Other biosignatures may be less conclusive in determining whether a planet hosts complex life. Infrared reflectance in Earth's spectrum known as the photosynthetic "red edge" is largely due to the presence of complex life such as terrestrial plants, moss, and lichens (Schwieterman et al. 2018). This might suggest that photosynthetic signatures may be less prevalent for planets orbiting low-mass stars, although a red edge itself is not unique to complex life. The development of complex life may also require a sufficient abundance of atmospheric oxygen (Catling et al. 2005) and could suggest that planets orbiting F-, G-, and early K-dwarf stars may be more likely to show signs of a strongly oxygenated atmosphere; however, oxygen itself is an insufficient condition to assume the presence of complex life. In general, technosignatures provide the least ambiguous observable method for testing PET and inferring the presence of complex life.

Molecular clocks provide a second method for testing the PET hypothesis. Long-term experiments with multiple lineages of *Escherichia coli* have been cultured for sixty thousand generations to explicitly simulate the dynamics of evolutionary selection over twenty-five years (Lenski 2017). Such experiments even show evidence of a relatively fixed rate of beneficial mutations— a constant molecular clock—over time (Barrick et al. 2009). Results such as these have suggested the possibility that the average rate of the molecular clock has remained approximately constant across Earth's history (Bromham and Penny 2003). The mean rate of the molecular clock depends upon the ultraviolet flux from the host star, so the PET hypothesis predicts that any application of a constant-rate molecular clock should be a function of the incident spectral energy distribution. One way of exploring this possibility is to examine the molecular clock in relevant terrestrial analogs, such as subsurface environments with less access to free energy (Jones et al. 2018). If the molecular clock of populations in such low-energy environments differs substantially from that of surface life, then this would offer empirical support of the PET hypothesis. A more intensive approach would be to conduct studies in experimental evolution, with multiple lineages of photosynthetic eukaryotes sustained under artificial lighting that correspond to the spectral energy distribution of F-, G-, K-, and M-dwarf stars. Once the populations have adapted to their conditions, subsequent evolution over ten to a hundred thousand or more generations should reveal a slower tick of the molecular clock for populations under the light of low-mass stars.

## 5.    Critiques and Next Steps

The purpose of this paper is to provoke discussion and stimulate new ideas about expectations for complex life in the galaxy. From a qualitative perspective, the PET hypothesis as summarized by Eq. (7) presents a monotonic function relating the evolutionary timescale of complex life to the mass of the host star (similar to the argument advanced by Livio and Kopelman (1990)). This approach to PET assumes a linear relationship between the stellar spectral energy distribution and the free energy on an orbiting habitable planet, which is a gross oversimplification to say the least. The function obtained by Livio and Kopelman (1990) took an exponential form due to their assumptions regarding atmospheric attenuation, while the use of a best-fit line in this paper is merely a device of convenience. However, any functional expression for the mechanical and chemical free energy available to life will almost certainly be non-linear and possibly non-monotonic, even if a general dependence on stellar mass still remains. The quality of photons could also impact the evolution of complex life; for example, a planet that receives low entropy photons emitted by an F-dwarf star might be more favorable to life than a similar planet that receives higher entropy photons from an M-dwarf star. The rate of the molecular clock likewise may depend upon photon quality and other nonlinear processes. This paper does not intend to imply that the relationship between evolutionary timescales and stellar mass is necessarily simplistic or predicable but only that such a relationship may exist. The specific form of the PET hypothesis as articulated here is intended to serve as a starting point for thinking more critically about the stellar environments likely to harbor complex life.

**Acknowledgments**

Thanks to Ravi Kopparapu, Eric Wolf, Omer Markovitch, Armando Azua-Bustos, Brendan Mullan, Gina Riggio, Sanjoy Som, Eddie Schwieterman, and Carl Pilcher for thoughtful discussions. Thanks also to Norm Sleep and two anonymous reviewers for helpful critiques. The author gratefully acknowledges funding from the NASA Habitable Worlds program under award 80NSSC17K0741. Any opinions, findings, and conclusions or recommendations expressed in this material are those of the author and do not necessarily reflect the views of NASA.